\documentclass{article}
\usepackage[english]{babel}
\usepackage{authblk}
\usepackage{upgreek}
\DeclareUnicodeCharacter{2009}{\,}
\usepackage[a4paper,top=1.5cm,bottom=2cm,left=1.5cm,right=1.5cm,marginparwidth=1.75cm]{geometry}

\usepackage{amsmath}
\usepackage{graphicx}
\usepackage[colorlinks=true, allcolors=blue]{hyperref}

\title{All optical operation of a superconducting photonic interface}
\author[1]{Frederik Thiele}
\author[1]{Thomas Hummel}
\author[2]{Adam N. McCaughan}
\author[1]{Julian Brockmeier}
\author[1]{Maximilian Protte}
\author[1]{Victor Quiring}
\author[1]{Sebastian Lengeling}
\author[1]{Christof Eigner}
\author[1]{Christine Silberhorn}
\author[1]{Tim J. Bartley}
\affil[1]{Institute for Photonic Quantum Systems, Paderborn University, Warburger Str. 100, Paderborn, Germany}%
\affil[2]{National Institute of Standards and Technology, 325 Broadway, Boulder, CO, USA, 80305}%

\begin{document}
\maketitle

\begin{abstract}
    Advanced electro-optic processing combines electrical control with optical modulation and detection. For quantum photonic applications these processes must be carried out at the single photon level with high efficiency and low noise. Integrated quantum photonics has made great strides achieving single photon manipulation by combining key components on integrated chips which are operated by external driving electronics. Nevertheless, electrical interconnects between driving electronics and the electro-optic components, some of which require cryogenic operating conditions, can introduce parasitic effects. Here we show an all-optical interface which simultaneously delivers the operation power to, and extracts the measurement signal from, an advanced photonic circuit, namely, bias and readout of a superconducting nanowire single photon detector (SNSPD) on a single stage in a 1K cryostat. To do so, we supply all power for the single photon detector, output signal conditioning, and electro-optic readout using optical interconnects alone, thereby fully decoupling the cryogenic circuitry from the external environment. This removes the need to heatsink electrical connections, and potentially offers low-loss, high-bandwidth signal processing. This method opens the possibility to operate other advanced electrically decoupled photonic circuits such as optical control and readout of superconducting circuits, and feedforward for photonic quantum computing.
\end{abstract}
\twocolumn

\section{Introduction}
      \begin{figure}[h]
            \centering
            \includegraphics[width=\linewidth]{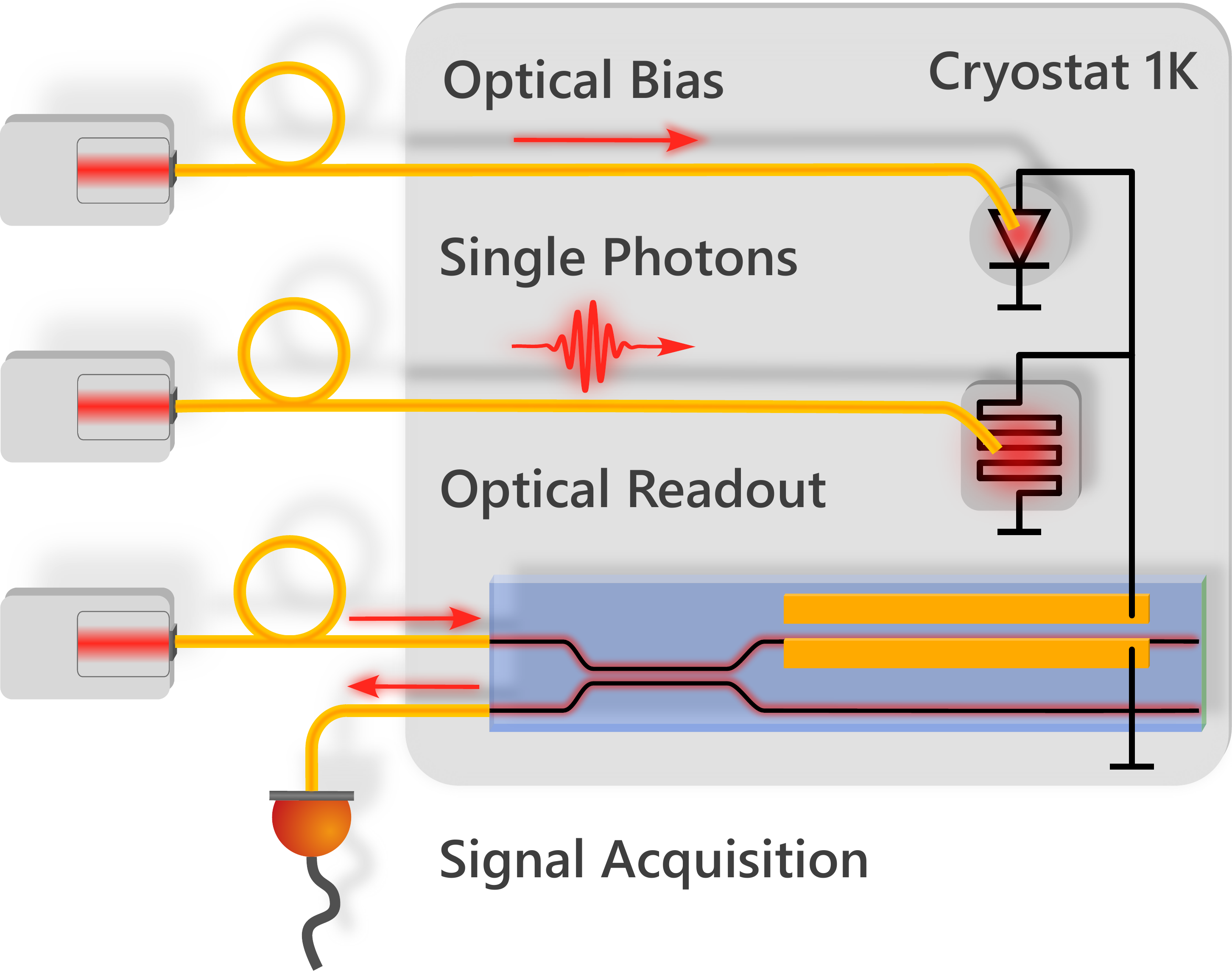}
            \caption{Layout of the all-optical operation of a Superconducting Nanowire Single Photon Detector (SNSPD). The cryogenic photodiode is illuminated and generates a photocurrent. When a photon impinges on the SNSPD it becomes resistive and voltage is created. To readout the detection signal, the voltage pulse is transmitted onto the cryogenic modulator. An intensity modulation is then readout at room temperature by a photodiode and timetagger.}
            \label{fig:sketch}
        \end{figure}
        Integrated quantum photonics offers great benefits for quantum information processing and communication~\cite{Bodanov2017,Wang2019a,Pelucchi2021}. With an increasing complexity of quantum photonic applications, a plurality of electro-optic components must be operated on a single chip~\cite{Moody2021,Kim2020a,EsmaeilZadeh2021}. These electro-optic components, such as modulators and single photon detectors, require additional ancillary electronic components for biasing, amplification, and signal transmission. Intermediate electric components will degrade the performance of the entire photonic setup by introducing noise, heatload, or bandwidth limitations~\cite{Krinner2019}. To circumvent this, it may be highly beneficial to replace the intermediate electronic connections with optical links, which electrically decouple the photonic processor and the driving electronics.     
        
        Superconducting nanowire single photon detectors\\(SNSPDs) are a key enabling technology for quantum optical applications due to their near unity detection efficiencies~\cite{Reddy2020,Chang2021}, low timing jitter~\cite{Korzh2020}, and low dark count rates~\cite{Hochberg2019}. Integrating these detectors in advanced photonic circuits is non-trivial since SNSPDs require operation temperatures below 4$\,\mathrm{K}$~\cite{Hadfield2016}. The cryogenic environment introduces additional challenges when interfacing the detectors with other electro-optic components. The output voltage of the SNSPD is around 1$\,\mathrm{mV}$~\cite{Hadfield2016} in the typical configuration with a 50$\,\mathrm{\Omega}$ shunt resistor. Therefore, additional electrical components are needed to amplify and transmit the SNSPD signal to the readout electronics. In the typical configuration the readout and bias electronics are outside the cryostat, requiring electrical interconnects between the cryogenic and room temperature environments. Replacing these electrical interconnects with electro-optic components and optical fibers enables electronic decoupling of the cryogenic electro-photonics from the external environment. An all-optical interconnect for superconducting photonics must therefore deliver the operation power and transmit signals to and from the decoupled circuit. 

        In recent years, cryogenic electro-optic modulation has been investigated for photonic circuits across a variety of platforms~\cite{Eltes2020,DeCea2020,Thiele2020,Lomonte2021,Gyger2021,Thiele2022a}. In particular, the electro-optic readout of superconducting single photon detectors both with and without intermediate amplifiers have been investigated~\cite{Youssefi2020,DeCea2020,Thiele2021}. In these applications, the click signal of an SNSPD detection event is delivered to an electro-optic modulator, modulating an optical throughput which is subsequently read out at room temperature. The operation voltage for intensity modulators have been reported to be in a range from 100$\,\mathrm{mV}$ to 10$\,\mathrm{V}$ at cryogenic temperatures~\cite{Eltes2020,DeCea2020,Lomonte2021,Thiele2022a}. Bridging the gap between the low amplitude output of the superconducting detector to these  voltages is non-trivial in a cryogenic environment. 

        In this paper, we present an alternative method to generate larger detection signals from an SNSPD, which drives an electro-optic modulator directly, to achieve an all-optical readout. The SNSPD works on the principle that a bias current is converted to a voltage signal through a resistive load. This resistive load is created by an impinging photon which breaks the superconductive state in the nanowire, followed by Joule heating creating a so-called hotspot. In a typical operation, a shunt resistor is introduced to redirect the bias current following a detection event. This prevents further Joule heating, allowing the nanowire to dissipate the heat from the hotspot and return to the superconductive state. In our configuration, shown in Fig.~\ref{fig:sketch}, we omit the shunt resistor allowing the resistive hotspot to grow to larger resistances due to self-heating. This method can generate resistances in the order of a few tens of$\,\mathrm{k\Omega}$ such that a signal voltage of 30$\,\mathrm{mV}$ is created. This is a significant increase compared to the 1$\,\mathrm{mV}$ signal from the conventional method. Nevertheless, in our configuration, the SNSPD remains ``latched'' in its normally resistive state, {i.e.} the hotspot does not have the ability to reset itself. Indeed, the latching dynamics and resulting output voltage of a nanowire is a largely unexploited effect~\cite{McCaughan2019,Baghdadi2020a}. 
        A key aspect of our method is the ability to actively reset the hotspot, ideally with a cryogenic current source. We achieve this by modulating the bias current optically with a current generated by a photodiode~\cite{Thiele2022c}. We have previously shown that this method shows no significant deviation from conventional biasing~\cite{Lecocq2021,Thiele2022c}. Thus the photodiode power delivery approach combines the key aspects of providing a bias current for the SNSPD and generating the electrical driving power of the modulator. This versatility in generating the SNSPD bias, sustaining the hotspot, and supplying the supply for the electro-optical modulation enables all-optical operation of the SNSPD. 
        
  \section{Results}
    \begin{figure*}
        \centering
        \includegraphics[width=\textwidth]{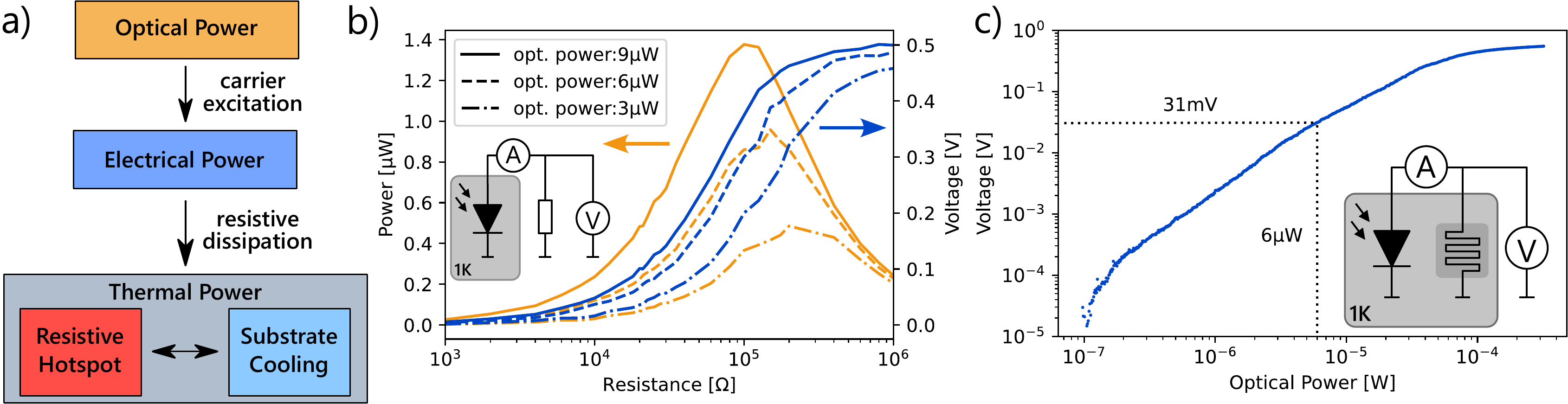}
        \caption{a) The optical input power is converted by the photodiode to an electrical power which is then converted to heat by Joule heating. b) Voltage and power output of the cryogenic photodiode under different illumination powers and load resistances. The photodiode is illuminated with a cw-laser at 1530$\,\mathrm{nm}$ and operated at 1$\,\mathrm{K}$. c) Voltage output of the combined operation of the SNSPD and bias photodiode, under different illumination powers to the bias photodiode at 1530$\,\mathrm{nm}$.}
        \label{fig:IVcryoPD}
    \end{figure*}


        Once a photon impinges on the nanowire, a hotspot is formed and the electrical power is converted to heat by the resistive nanowire through Joule heating, as depicted schematically in Fig.~\ref{fig:IVcryoPD}~a). This Joule heating results in heating of adjacent superconductive regions, increasing the hotspot size. The resistive region of the nanowire grows until the electrical power provided is equal to the power dissipated due to the temperature difference between the nanowire and the substrate~\cite{Yang2007a,Berggren2018}. At the power equilibrium the maximum output voltage is reached, generated by the current flowing trough the resistive\\nanowire. The output voltage of the latched SNSPD is therefore dependent on the power dissipation of the SNSPD and the electrical power provided by the photodiode. We investigate this by characterising the power conversion capabilities and the output voltage of the photodiode in combination with the SNSPD.     
     
        The optical to electrical conversion efficiency of the photodiode depends on the load resistance, in our case the SNSPD. The SNSPD resistance can vary from no resistance when superconducting to a normal resistance of 5.5$\,\mathrm{M\Omega}$ when the entire nanowire is normally resistive. To determine the generated voltages and dissipated powers, we cooled the photodiode down to 1$\,\mathrm{K}$ and attached different load resistors at room temperature according to the inset in Fig.~\ref{fig:IVcryoPD}~b). The photodiode is illuminated with an optical input power of about 6$\,\upmu\mathrm{W}$. At this input power, a nominal bias current of 4$\,\upmu\mathrm{A}$ for the SNSPD is generated in shunted operation, given that the responsivity of the photodiode is approximately 0.65$\,\mathrm{A/W}$~\cite{Thiele2022c}. In our characterisation, we keep the input power stable and vary the load resistor in a range from 10$\,\mathrm{\Omega}$ to 1$\,\mathrm{M\Omega}$, while measuring the current through and voltage over the resistor. As a result, we generate increasing output voltages by increasing the load resistors, until the output voltage saturates at approximately 500$\,\mathrm{mV}$, as can be seen in Fig.~\ref{fig:IVcryoPD}~b). The generated electrical power reaches a maximum power point at a resistance around 100$\,\mathrm{k\Omega}$, before reducing again for higher loads due to a reduction in the generated current. This maximum power point shifts to lower resistances when the optical power on the photodiode is increased. This load characterisation shows that the cryogenic photodiode can provide a bias current in the shunted operation when the nanowire is superconductive and a voltage beyond 500$\,\mathrm{mV}$ when a resistive hotspot is created.
     
        The hotspot in the nanowire is not a static resistor and is expected to change in size depending on the electrical supply power. Therefore, the output voltage will depend on the optical input power of the photodiode. However, we cannot increase the input power to the photodiode indefinitely to generate a maximal optical response because the SNSPD must be operated at a nominal bias current of 4$\,\upmu\mathrm{A}$. Therefore, we need to determine the voltage at this operation point when combining the SNSPD directly with the photodiode. To characterize the devices before the all-optical demonstration, we combined both devices on a single stage in the cryostat at 1K and read out the output voltage at different input powers the photodiode, as shown in the inset of Fig.~\ref{fig:IVcryoPD}~c). Both devices are connected with a coax cable to measure the voltages at room temperature. Fig.~\ref{fig:IVcryoPD}~c) shows that voltages with the photodiode and SNSPD are generated and saturate above approximately 550$\,\mathrm{mV}$ when we sweep the input power from 0.1$\,\upmu\mathrm{W}$ to 700$\,\upmu\mathrm{W}$. In the all-optical operation of the SNSPD, we provide an input power of 6$\,\upmu\mathrm{W}$ to generate the nominal bias current for the SNSPD. At this power level a output voltage of 31$\,\mathrm{mV}$ is reached, generated by a 40$\,\mathrm{k\Omega}$ nanowwire resistance. This output voltage is a significant increase in relation to a click signal of below 1$\,\mathrm{mV}$ in the conventional method.


        \begin{figure}
            \centering
            \includegraphics[width=\linewidth]{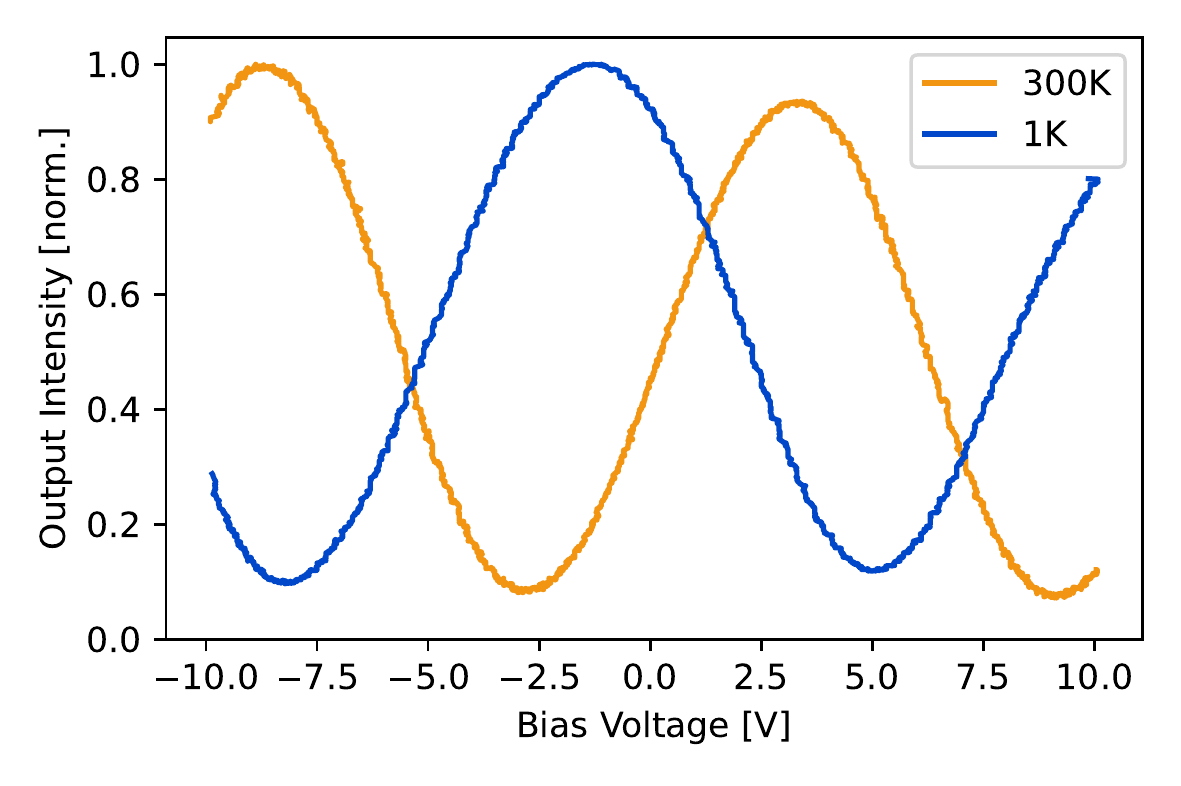}
            \caption{Voltage sweeps of the modulators at 1530$\,\mathrm{nm}$. The measured intensities are normalised to the maximal out power per sweep. The $V_{\pi}$ voltage voltages are 5.9$\,\mathrm{V}$ at 1$\,\mathrm{K}$ and 6.6$\,\mathrm{V}$ at 300$\,\mathrm{K}$. The $V_{\pi}$ voltage is acquired by fitting a sin-function to the acquired data.}
            \label{fig:CompVpi}
        \end{figure}
            
        The readout of the generated click signal is realised by modulating the intensity of an electro-optic modulator operated at cryogenic temperatures. We choose a titanium in-diffused lithium niobate electro-optic modulator because of its proven operation at cryogenic temperatures~\cite{Thiele2020,Thiele2022a}. Furthermore, we can achieve fibre-to-fibre optical coupling up to 43\% with single mode fibres even at 1$\,\mathrm{K}$~\cite{Thiele2020}. We implement an integrated Michelson interferometer consisting of a directional coupler, a reflective endface coating, and electrodes placed on one channel for the modulation. Optical access is realised with a dual-core ferrule housing standard single mode fibres, such that one fibre transmits the light into the waveguide and the second fibre receives the reflected light. We achieve a fibre-to-fibre efficiency of 27\% when operating at 1\,K. We characterised the electro-optic modulator by acquiring the intensity of the output by sweeping the voltage, as shown in Fig.~\ref{fig:CompVpi}. The modulation voltage required to switch the intensity from maximum to minimum is extracted to be 6.6$\,\mathrm{V}$ at 1$\,\mathrm{K}$, at an operation wavelength of 1530$\,\mathrm{nm}$. More details on the fabrication, waveguide characterisation and fibre access are given in the Methods section.

        \begin{figure}
            \centering
            \includegraphics[width=\linewidth]{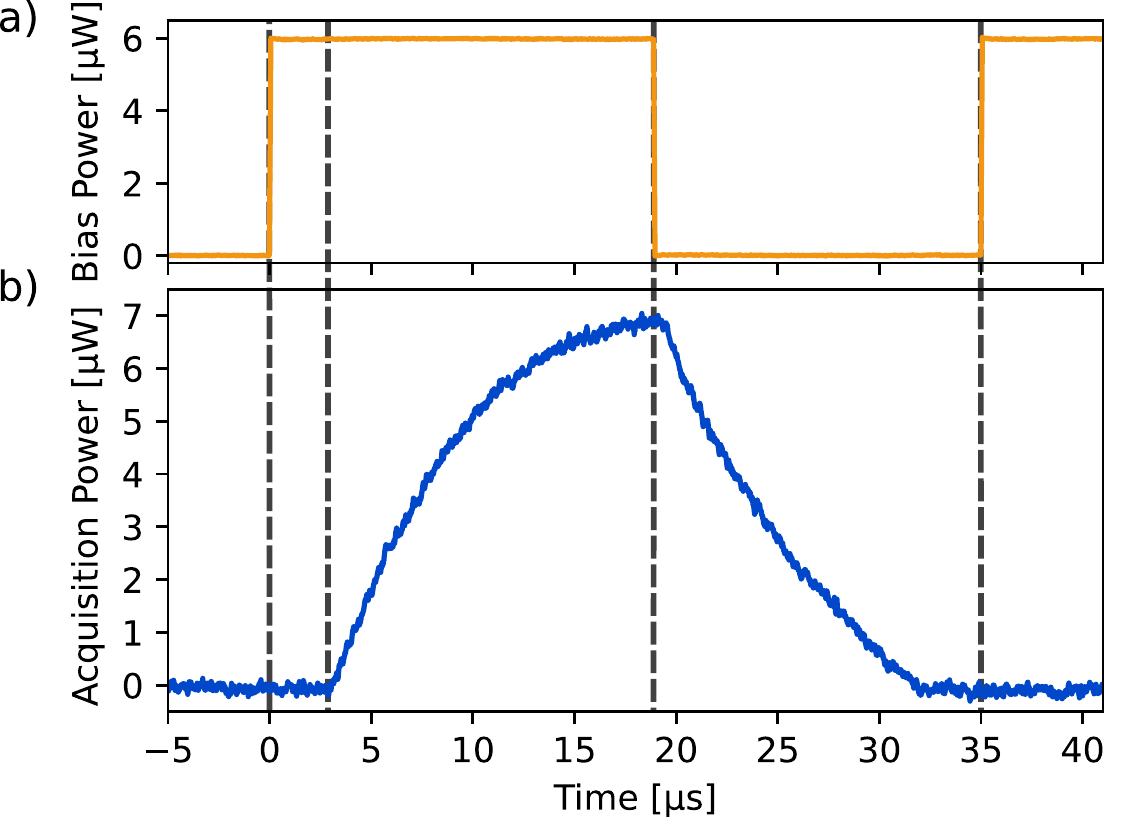}
            \caption{a) Optical modulation of the bias power for the cryogenic photodiode to operate the SNSPD. The modulation period is 35$\,\upmu\mathrm{s}$. b) The optical response of the readout photodiode after the electro-optic modulator placed at room temperature. A photon impinges on the SNSPD at 2.9$\,\upmu\mathrm{s}$, a voltage is generated and switches the electro-optic modulator. To reset the SNSPD, the optical power to the cryogenic photodiode is switched off after 18.9$\,\upmu\mathrm{s}$. The nanowire becomes superconducting such that the generated voltage over nanowire resistance is reduced to zero.}
            \label{fig:Trace}
        \end{figure}
        
        We combine the operation of an opto-electronic bias of the SNSPD with an electro-optic readout. To do so, the photodiode, SNSPD, and lithium niobate modulator are connected on a single cold stage of the cryostat and cooled down to a base temperature of 1$\,\mathrm{K}$. For the operation of the SNSPD a 6$\,\upmu\mathrm{W}$ cw-laser is externally on-off pulsed with a duty cycle of 35$\,\upmu\mathrm{s}$, as illustrated in Fig.~\ref{fig:Trace}~a). A detection event occurs 2.9$\,\upmu\mathrm{s}$ after the illumination of the photodiode. As a result, a hotspot starts to grow, resulting in an increasing voltage delivered to the modulator. To read out the generated detection signal, light is transmitted through the modulator and the optical response is measured at room temperature. A typical measurement trace is shown in Fig.~\ref{fig:Trace}b). The resulting optical response signal has a 90\% rise time of 11$\,\upmu\mathrm{s}$. Once a photon is detected the superconductor becomes resistive and no subsequent photons can be detected, because the current through the nanowire is strongly reduced. The SNSPD is reset by switching off the light to the biasing photodiode after 18.9$\,\upmu\mathrm{s}$, which results in an optical response with a fall time of 11$\,\upmu\mathrm{s}$. To maximise the optical response, we tuned the operation wavelength of the modulator to 1530$\,\mathrm{nm}$ and optimised the input polarisation with a fibre polarisation controller while a power of 3.5$\,\mathrm{mW}$ is transmitted through the modulator. The optical response is read out with a photodiode at room temperature, resulting in a signal with an amplitude of 6.9$\,\mathrm{\upmu W}$. This shows the core principle that we can all-optically bias an SNSPD and read out the detection signal.

\begin{figure}
            \centering
            \includegraphics[width=\linewidth]{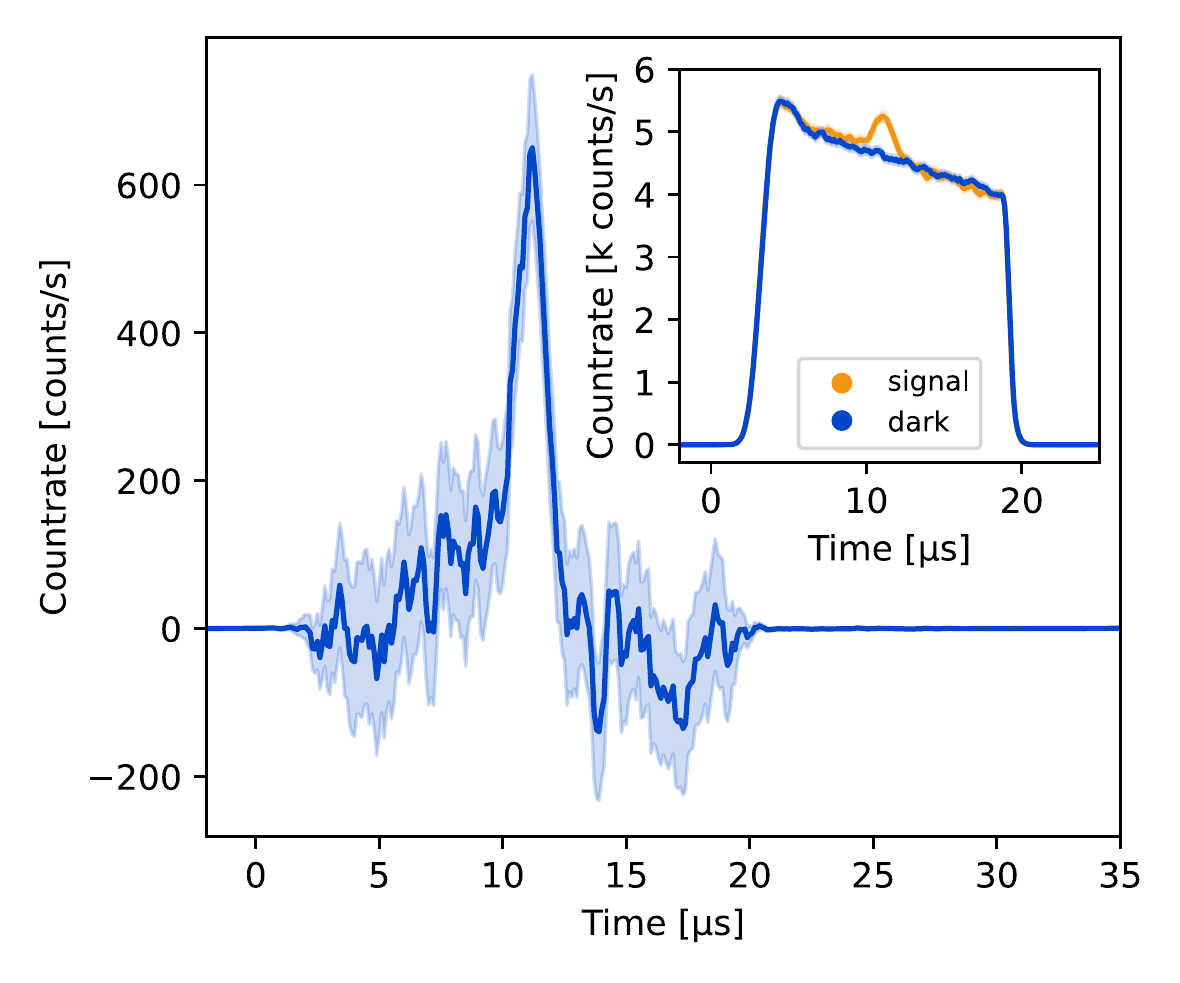}
            \caption{Histogram of the acquired countrate of the all-optical SNSPD operation. This is the difference in the countrate of the signal photons and measured dark counts. The error bars are determined by the counting errors of the count rates. The inset shows the total countrate of the measurement with and without a single photon input to the SNSPD. The mean photon number per pulse is 1.17$\pm0.06$.}
            \label{fig:histdiff}
        \end{figure}

        As a next step, we characterise the single photon response of the superconducting detector. To do so, we operated the SNSPD with our all-optical biasing method and acquired the countrate of the SNSPD with a single photon level input. This comprises attenuated laser pulses with a mean photon number of 1.17$\pm0.06$ photons per pulse at 1545$\,\mathrm{nm}$. This optical input is then pulsed synchronously with the optical bias operated with a 35$\,\upmu\mathrm{s}$ on-off period. The timing delay of the generated click signals are then recorded in relation to the on-off signal. The readout is realised by the modulator. To generate click signals from the optical response signal, the room temperature readout photodiode is replaced by a Small Formfactor Pluggable module (SFP) with a specified pulse sensitivity of -25$\,\mathrm{dBm}$. As a result, we can minimize the input power to the modulator to 68$\,\upmu\mathrm{W}$ while the click signals are still detected. The click signals with a single photon input are then acquired with a time tagger and displayed as a histogram in Fig.~\ref{fig:histdiff}. During the measurement, signal photons as well as scattered light from the electro-optic modulator are detected. To extract the single photon events we performed a background measurement without a single photon input. By subtracting this background from the results with a single-photon input, a clear peak in the countrate at a delay of 11.3$\,\upmu\mathrm{s}$ with a full width half maximum of 1.5$\,\upmu\mathrm{s}$ can be seen, as shown in Fig.~\ref{fig:histdiff}. The countrate in this histogram is negative after the main peak because the previous signal photons decrease the detection probability for subsequent background photons due to the SNSPD latching. The peak shows clearly that the SNSPD is sensitive to a single photon input in our all-optical operation method. 

\section{Discussion}
    The proof-of-principle devices presented here show promise for interfacing superconducting photonics in new performance regimes and application spaces, however there are still some non-idealities that can be improved in future work. A strong background of dark counts is present, mainly generated by scattered light introduced by the optical bias and readout of the SNSPD. Due to the lack of a self-reset, the detector clicks on the first photon that it measures, and none after that. Therefore, reducing the number of premature latching clicks from noise is necessary. In the present configuration, the components are mounted in the cryostat without intermediate shielding, the application of which would reduce the dark count rates significantly. Furthermore, one could exploit differences in the spectral sensitivity between the bias photodiode and the SNSPD to further reduce noise.
    
    The SNSPD was operated with a repetition rate around 28$\,\mathrm{kHz}$, limited by the rise and fall time of the click signals. We attribute this to the charging dynamics of the capacitance of the electro-optic modulator and the growth of the resistive hotspot. Faster rise times can be achieved by introducing a parallel resistance to reduce the hotspot size. Introducing a parallel resistance also reduces the output voltage and hence the intensity modulation. Therefore, there exists a trade-off between the output voltage and the resulting rise time.
    
    This bias and readout method has no effect on the internal detection efficiency of the SNSPD. We have previously shown that an optical bias and a conventional bias method achieve the same detection efficiencies~\cite{Thiele2022c}. Nevertheless, the lack of a self reset means that "true" counts may be lost if the detector has previously been triggered by a dark count, therefore reducing the noise also plays an important role in increasing the system efficiency. This could be further optimised by careful synchronisation of bias and readout light sources with respect to the expected arrival time of the single photons.
    
    Independent of optical biasing and readout, exploiting the latched state of an SNSPD is an important method to step up the click signal. For example, amplitudes of 10$\,\mathrm{mV}$ could be used to drive a Schmitt-trigger, which can enable feedforward modulation with electronic amplifiers. With this technique, the initial click signal is significantly increased, such that cryogenic low noise amplifiers can be avoided. 
    The key to this method is to integrate a current source which can reset the latched SNSPD after the detection. 
    
    By connecting the output of the detector to a modulator, this method is an important step towards all-optical feedforward modulation~\cite{Prevedel2007} at cryogenic temperatures, which is integral to quantum photonic one-way computing~\cite{Raussendorf2001}. This can be achieved by a direct matching of the generated output voltage of the SNSPD signal and the switching voltage of the modulator. In this technique, a photon can be detected by the single photon detector switching the electro-optic modulator from one state to the other. Future work is needed to match the detector output and modulation voltages at cryogenic temperatures. In addition, the rise time of the click signal must be improved to minimise optical delays when using feedforward modulation with low latency processes.     
    


    Since optical fibres inherently have a lower thermal conductivity than coaxial cables, this can significantly reduce the heatload on the cryostat~\cite{DeCea2020,Lecocq2021}. The passive heatload of coaxial cables can be mitigated by thermal anchoring in the cryostat but are a major contributor to the thermal load on a cryostat~\cite{Krinner2019}. In our operation of the SNSPD, three independent optical fibres are directly connected from room temperature to the photodiode and electro-optical modulator without the need of intermediate thermal anchoring. The active heatload of these electro-optic devices are 6$\,\upmu\mathrm{W}$ for the photodiode and 68$\,\upmu\mathrm{W}$ for the electro-optic modulator. The active heatload of the modulator can be reduced further to a few $\upmu\mathrm{W}$ by reducing the modulation voltage $V_{\pi}$ and increasing the detection signal voltage, such that the full throughput intensity is modulated. In addition, wavelength-division multiplexing can also be used for the bias and readout to operate multiple SNSPDs in parallel to reduce the passive heatload even further~\cite{Lecocq2021}.      

    In summary, we have realised an all-optical interface for quantum photonic applications. The interface provides the operation power for a superconducting single photon detector via a cryogenic photodiode. In addition, the detection signals are readout optically via an electro-optic modulator at cryogenic temperatures. The all-optical operation shows promising techniques for the combined operation of superconducting electronics and photonics circuits, which are electrically isolated from their driving circuitry. Increasing the signal amplitudes and modulation capabilities of the opto-electronic components enables further applications such as feed-forward. Furthermore, the all-optical operation of the SNSPD achieves a low power operation of the SNSPD by providing only a total operation power of 75$\,\mathrm{\upmu W}$.    

\section{Acknowledgements}
    This work was supported by the Bundesministerium für Bildung und Forschung (Grant No. 13N14911) and co-funded by the European Union (ERC, QuESADILLA,\\101042399). Views and opinions expressed are however those of the author(s) only and do not necessarily reflect those of the European Union or the European Research Council. Neither the European Union nor the granting authority can be held responsible for them. We thank Varun Verma (NIST) for providing the superconducting films for the Superconducting Nanowire Single Photon Detectors.
    
\bibliographystyle{ieeetr}


\section{Methods}

\subsection{Michelson interferometer}
\begin{figure}
            \centering
            \includegraphics[width=\linewidth]{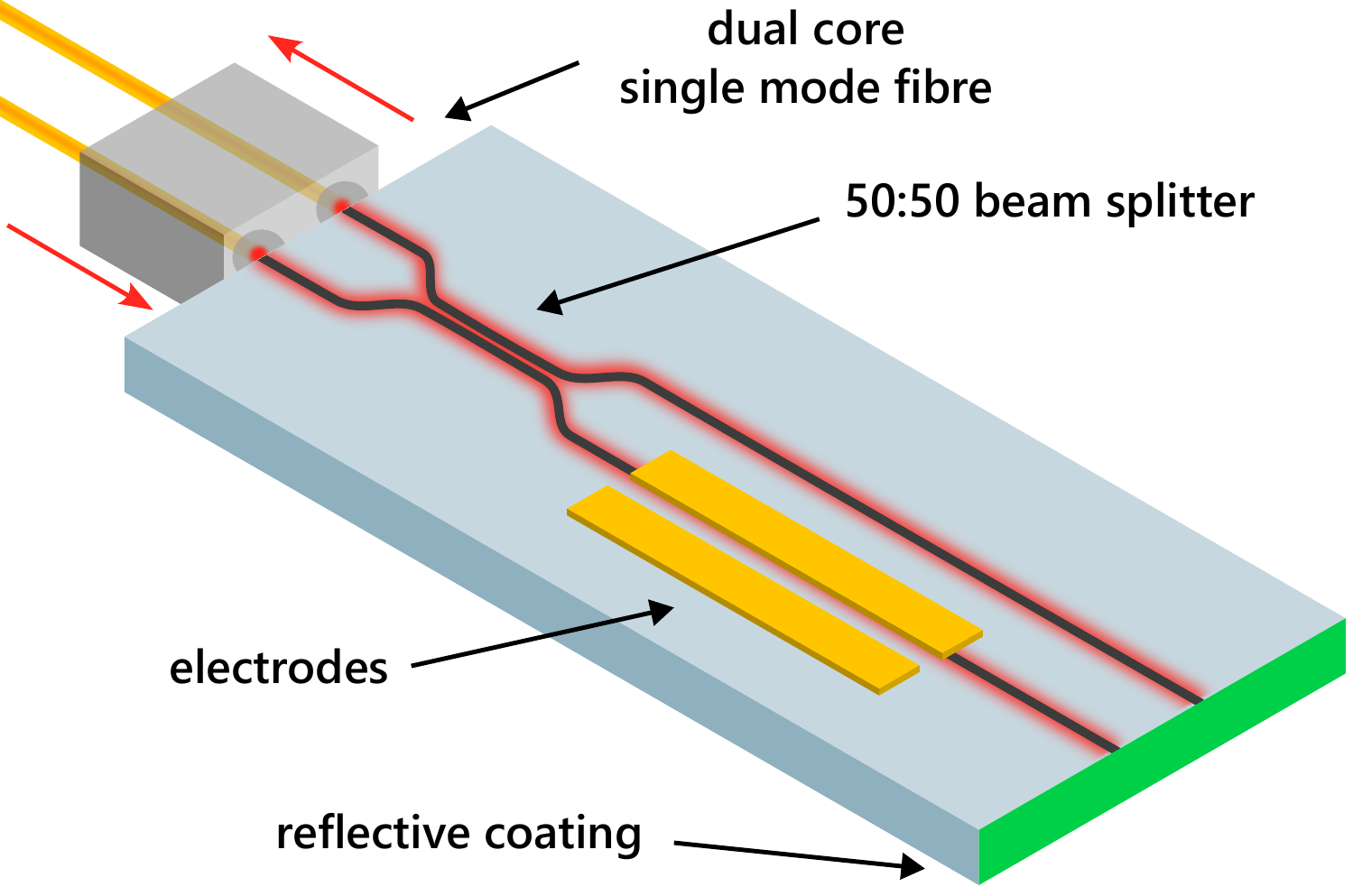}
            \caption{Layout of the Michelson-interferometer realised in a photonic circuit. A dual core single mode fibre pigtail couples light into the waveguides and returns the reflected light for the readout. These waveguides are fabricated by titanium indiffusion in z-cut lithium niobate. An integrated beam splitter splits the input light into two paths which are then reflected at the endface to interfere again at the beamsplitter. Electrodes on the surface of one beam arm introduce a phase difference when a voltage is applied.}
            \label{fig:OptAccess}
        \end{figure}

        In this setup, the cryogenic readout of the detection signals is realised through an electro-optic modulation. We choose titanium indiffused lithium niobate as a material platform to realise an integrated intensity modulator. In this platform we can use the electro-optic effect even at cryogenic temperatures as well as achieve a high coupling efficiency between a single mode fibre and the waveguide. The electro-optic modulation is realised in a Michelson-interferometer in which a beamsplitter, phase modulator and endface reflector are integrated with waveguides, as illustrated in Fig.~\ref{fig:OptAccess}. Light is coupled into the chip and split into two paths by the 50:50 beamsplitter. On one arm of the beamsplitter light is modulated by the electro-optic phase shifter. Both beams are then reflected at the endface of the chips with an endface coating. The reflected light is propagating in the reverse direction and interfere on the beamsplitter again. As a result, an intensity modulation can be read out when the voltage on the electro-optic modulator is varied because the relative phase between the beams is changed. The Michelson-interferometer layout is chosen to reduce the required voltage to introduce the phase shift because the phase shift is accumulated in both the forward and backward propagation through the modulator.
        The waveguides are realised in this lithium niobate platform by a titanium in-diffusion process. To do so, titanium is deposited on top of the sample and patterned with laser photo-lithography and wet-etching. A resulting 7 $\mathrm{\upmu m}$ wide titanium stripe is then diffused into the sample in an oven. The in-diffused titanium increases the reflective index such that wave guiding is achieved.  The optical loss in the waveguides are characterised with the Fabry-Pérot method with a straight waveguides next to the Michelson-interferometer before the reflective coating is applied~\cite{Regener1985}. The resulting losses are about 0.1dB/cm at a wavelength of 1550$\,\mathrm{nm}$ in the TE polarisation. 

        The integrated 50:50 beamsplitter is realised by an evanescent coupler consisting of two parallel waveguides, as illustrated in Fig.~\ref{fig:OptAccess}. The waveguides separation distance is reduced from 127$\,\mathrm{\upmu m}$ to 13$\,\mathrm{\upmu m}$ by S-bends to interfere on 8.3$\,\mathrm{mm}$ long distance and then separated again. This coupling length is chosen such that a splitting ratio of 50:50 is reached between the optical path in both polarisations TE and TM. The total length of the sample is 56$\,\mathrm{mm}$, including the beam splitter and modulator. 
    
        A phase shift in the Michelson-interferometer is realised by an electro-optic modulator. An electric field in the waveguide is introduced when a voltage is applied to electrodes on the surface of the sample. This electric field introduces a refractive index change in the waveguide, which induces a relative phase shift between the beamsplitter arms, as it can be seen in Fig.~\ref{fig:OptAccess}. The modulator's electrodes are placed behind the beam splitter with respect to the input facet. The electrode pair is positioned parallel to the waveguide such that the inner edge of one electrode overlaps the waveguide to induce a vertical electric field. The second electrode is placed next to the waveguide and is separated distance of 9$\,\mathrm{\upmu m}$ from the first electrode. The electrode pair has a length of 20$\,\mathrm{mm}$ and overlaps the waveguide over the full length. The electrode stack consisting of a 400$\,\mathrm{nm}$ $SiO_2$ bottom layer, 10$\,\mathrm{nm}$ titanium adhesion-layer and 100$\,\mathrm{nm}$ gold layer which is selectively deposited with photo-lithography and lift-off process. 
    
        The reflector in the Michelson setup is realised by an endface coating on the endfacet at opposite side to the fibre input. To do so, the endface is polished using a CMP method. Afterwards a dielectric coating is deposited by evaporating titanium oxide and silicon oxide using an oxigen ion assisted electron beam evaporation method~\cite{Stefszky2021}. The dielectric-stack is designed such that a high reflectivity is reached in a range from 1500$\,\mathrm{nm}$ to 1600$\,\mathrm{nm}$. The reflectivity is characterised by measuring the transmission through a reference lithium niobate sample with the same dielectric stack. A reflectivity of 96\% is achieved at the operation wavelength of our experiment, namely 1530$\,\mathrm{nm}$.      
    
        The integrated Michelson-interferometer achieves the strongest intensity modulation when a splitting ratio of 50:50 is reached for the beamsplitter. To characterise the intensity modulation we coupled light into the waveguide and swept the bias voltage at the electrodes. A relative phase difference of $\pi$ between the modulated and unmodulated path is needed to switch the intensity from a minimum to a maximum. The bias voltage dependent intensity follows a sin-function such that we can fit this function to extract the $V_\pi$-voltage, compare Fig.~\ref{fig:CompVpi}. This modulation voltage is characterised to be 5.9$\,\mathrm{V}$ at room temperature and increases to 6.6$\,\mathrm{V}$ when cooled down to an operation temperature of 1$\,\mathrm{K}$. The increase in the $V_\pi$-voltage is due to the temperature-dependent electro-optic coefficient which has been investigated previously \cite{Herzog2008,Thiele2022a}. In summary, our electro-optic modulator reaches a modulation strength of 23.6$\,\mathrm{Vcm}$ at an operation wavelength of 1530$\,\mathrm{nm}$.    
    
    \subsection{Optical access}
        The optical access to our photonic circuit is of high importance in out all-optical operation. The link has to achieve a high coupling efficiency to our waveguides even at cryogenic temperatures. Additionally, the circuit requires the access for the input and the reflected output port, as illustrated in Fig.~\ref{fig:OptAccess}. Therefore, a fibre ferrule with two single mode fibres are used with a core separation of 127$\,\mathrm{\upmu m}$ which is equal to the waveguide separation at the endface. To attach the fibre ferrule to the lithium niobate sample, the fibre position is optimised by maximising the reflected power through the fibres and waveguides. Afterwards, the fibres position is permanently fixed by a UV-adhesive (NOA81) at the fibre-to-sample interface. The sample and attached fibres are then placed on a mounting block for mechanical stability such that the lithium niobate is fixed to the surface with vacuum grease (Apiezon N) and the dual core single mode fibre is fixed with UV-cured adhesive to the holder.
        We achieve a fibre-to-fibre efficiency of about 45\% at 1530nm at room temperature which is reduced to 27\% at cryogenic temperatures. The change in the coupling efficiency can be mainly attributed to mechanical changes in the adhesive bond during the cooling process. The supported optical mode in our waveguide have a large mode overlap with standard single mode fibres in the TM polarisation of 85\% and with the TE polarisation of 92\%~\cite{Montaut2017d}. We can expect an overall coupling efficiency between the waveguides and single mode fibre of 51\% given a mode overlap of 85\%, a linear loss of 0.1dB/cm, a coating reflectivity of 96\% and a fibre to waveguide reflectivity of 97\%.

     \subsection{Cryogenic Photodiode}
        The cryogenic photodiode is the power source to operate the electro-optical circuits. This photodiode must be able to convert an optical input power to an electrical power at cryogenic temperatures. The key to achieve this is that the diode material must still be responsive at cryogenic temperatures~\cite{Zhang1997,Bardalen2018a, Thiele2022c}. We used the same set of off-the-shelf photodiodes as in previous works~\cite{Thiele2022c}. The optical access is achieved by attaching a single mode fibre above the photodiode with a custom holder, achieving a responsivity of 0.65$\,\mathrm{A/W}$.

    \subsection{Optical Acquisition}
        The optical response of the modualtor is read out with a photodiode at room temperature, resulting in a signal with an amplitude of 6.9$\,\mathrm{\upmu W}$, given that the photodiode has a responsivity of 0.4\,$\mathrm{mV/\upmu W}$ and the readout amplitude is 2.76$\,\mathrm{mV}$, as shown in Fig.~\ref{fig:Trace}. We expect an optical response of about 7.0$\,\mathrm{\upmu W}$, given a click signal amplitude of 31$\,\mathrm{mV}$, the $\mathrm{V_\pi}$ of the modulator is 6.6$\,\mathrm{V}$ and a fibre-to-fibre efficiency of the modulator of 27\%. 
    
        Based on these results, we seek to reduce the optical throughput of the modulator to reduce both heatload and scatter. This depends on the minimal acquisition power of the detectors used to acquire the click signals. 
        Small Formfactor Pluggable modules (SFP) are ideal candidates to measure small changes in intensity. 
        The minimal nominal input pulse power of the SFP module is at 1550nm about -25$\,\mathrm{dBm}$ which will generate an electrical output pulse of above 100$\,\mathrm{mV}$ (Finisar FWLF-1519-7D-59). In our experiment we reduced the input power to the modulator to about 68$\,\mathrm{\upmu W}$ ($\sim$-12$\,\mathrm{dBm}$) such that the click signal has an amplitude of about 250$\,\mathrm{nW}$($\sim$-36$\,\mathrm{dBm}$). When acquiring these signals with the SFP-module, the generated output signal of the module was reduced to about 20$\,\mathrm{mV}$. Additional high frequency noise in the output signal is also introduced at these low input powers which we reduced with a 1$\,\mathrm{MHz}$ low-pass filter. This added noise is negligible at higher input powers and higher signal count rates. During the histogram acquisition in the all-optical operation of the SNSPD the time tagger did not trigger on the added noise since the countrate is zero when the cryogenic photodiode not illuminated.
        
    \subsection{SNSPD}
        The Superconducting Nanowire Single Photon Detector (SNSPD) is realised with a tungsten silicide nanowire (WSi) thin film. The nanowire is fabricated on the a silicon base substrate as it is described in \cite{Krapick2017}. The film has a normal resistance at 10K of 5.5$\,\mathrm{M \Omega}$. The bias current dependent system detection efficiency was previously characterised in Thiele et al. \cite{Thiele2022c}. We operated the SNSPD with an input power of 6$\,\mathrm{\upmu W}$ such that the SNSPD is biased at a nominal bias current of 4$\,\mathrm{\upmu A}$. The detection of the SNSPD is about 83\% $\pm 5\%$.

\end{document}